\documentclass[a4paper,11pt]{article}
\pdfoutput=1

\usepackage[T1]{fontenc}
\usepackage{lmodern}
\usepackage{amsmath}
\usepackage{amsfonts}
\usepackage{amssymb}
\usepackage{mathrsfs}
\usepackage{physics}
\usepackage{graphicx}
\usepackage{caption}
\usepackage{subcaption}
\usepackage{jheppub}
\usepackage{enumitem}
\usepackage{framed,float}
\usepackage{booktabs}
\usepackage{array}
\usepackage{soul}
\usepackage{multirow}
\usepackage{comment}
\usepackage{braket}
\usepackage{todonotes}
\usepackage{youngtab}
\usepackage{tikz}
\usetikzlibrary{calc,arrows,decorations.markings}

\usepackage{color}

\preprint{}
\title{\boldmath Proof of $A_{n}$ AGT conjecture at $\beta=1$}

\author[a]{Qing-Jie Yuan,}
\author[a]{Shao-Ping Hu,}
\author[a]{Zi-Hao Huang}
\author[a,b,c,1]{and Kilar Zhang\note{Corresponding author.}}
\affiliation[a]{Department of Physics and Institute for Quantum Science and Technology, Shanghai University, Shanghai 200444, China}
\affiliation[b]{Shanghai Key Lab for Astrophysics, Shanghai 200234, China}
\affiliation[c]{Shanghai Key Laboratory of High Temperature Superconductors, Shanghai 200444, China}
\emailAdd{wolfgangyuan@shu.edu.cn}
\emailAdd{279579878@shu.edu.cn}
\emailAdd{huangzihao@shu.edu.cn}
\emailAdd{kilar@shu.edu.cn}

\abstract{AGT conjecture reveals a connection between 4D $\mathcal{N}=2$ gauge theory and 2D conformal field theory. Though some special instances have been proven, others remain elusive and the attempts on its full proof never stop.  When the $\Omega$ background parameters satisfy $-\epsilon_1/\epsilon_2\equiv \beta =1$, the story can be simplified a bit.  A proof of the correspondence
in the case of $A_{1}$ gauge group was given in 2010 by Mironov et al., while the $A_{n}$ extension is verified by Matsuo and Zhang in 2011,  with an assumption on the Selberg integral of $n+1$ Schur polynomials. Then in 2020,  Albion et al.  obtained the rigorous result of this formula.  In this paper, we show that the conjecture on the Selberg integral of Schur polynomials  is formally equivalent to their result, after applying a more complicated complex contour, thus leading to the proof of the $A_{n}$ case at $\beta=1$.  To perform a double check, we also directly start from this formula,  and manage to show the identification between the two sides of AGT correspondence.
\\
\\
$${\textbf{In Memory of Professor Chaiho Rim}}$$}

\begin{document}

\newcommand{\hJ}{\hat{J}}
\newcommand{\hL}{\hat{L}}

\newcommand{\bs}{\backslash}
\newcommand{\sqz}{\mathbf{z}}
\newcommand{\leg}{\mathrm{Leg}}
\newcommand{\arm}{\mathrm{Arm}}
\newcommand{\rrangle}{\rangle\!\rangle}
\newcommand{\llangle}{\langle\!\langle}

\newcommand{\Int}{\int\limits}

\def\a{\alpha}
\def\b{\beta}
\def\c{\varepsilon}
\def\d{\delta}
\def\e{\epsilon}
\def\f{\phi}
\def\g{\gamma}
\def\h{\theta}
\def\k{\kappa}
\def\l{\lambda}
\def\m{\mu}
\def\n{\nu}
\def\p{\psi}
\def\q{\partial}
\def\r{\rho}
\def\s{\sigma}
\def\t{\tau}
\def\u{\upsilon}
\def\v{\varphi}
\def\w{\omega}
\def\x{\xi}
\def\y{\eta}
\def\z{\zeta}
\def\D{\Delta}
\def\G{\Gamma}
\def\H{\Theta}
\def\L{\Lambda}
\def\F{\Phi}
\def\P{\Psi}
\def\S{\Sigma}
\def\o{\over}

\def\bu{\bar{u}}
\def\bN{\bar{N}}
\def\bv{\bar{v}}
\def\bY{\bar{Y}}
\def\bx{\bar{x}}

\def\Zv{{Z}_\text{vect}}
\def\Zbf{{Z}_\text{bif}}
\def\Zf{{Z}_\text{fund}}
\def\bZbf{\bar{\mathcal{Z}}_\text{bif}}

\def\aY{|\vec{a},\vec{Y}\rangle}
\def\Ga{|G,\vec{a}\rangle}
\def\bGa{\langle G,\vec{a}|}
\def\CV{\mathcal{V}}
\def\tN{\tilde{N}}
\def\Zi{\mathcal{Z}_{\text{inst}}}
\def\nQ{Q}
\def\mf{m^{(f)}}

\def\cA{{\mathcal{A}}}
\def\cZ{{\mathcal{Z}}}
\def\cB{{\mathcal{B}}}
\def\cH{{\mathcal{H}}}
\def\cI{{\mathcal{I}}}
\def\cM{{\mathcal{M}}}
\def\cA{{\mathcal{N}}}
\def\cO{{\mathcal{O}}}
\def\cL{{\mathcal{L}}}
\def\cR{{\mathcal{R}}}

%from 2002.05637
\newcommand{\tar}[1]{t^{(#1)}}
\newcommand{\xar}[1]{x^{(#1)}}
\newcommand{\lar}[1]{\lambda^{(#1)}}
\newcommand{\Yar}[1]{Y^{(#1)}}
\newcommand{\Abs}[1]{\big\lvert#1\big\rvert}
\newcommand{\dup}{\mathrm{d}}
\newcommand{\iup}{\hspace{1pt}\mathrm{i}\hspace{1pt}}
%%%%%%%%
\def \be  {\begin{equation}}
\def \ee  {\end{equation}}
\def \ba {\begin{equation}\begin{aligned}}
\def \ea {\end{aligned}\end{equation}}
\def \bea  {\begin{eqnarray}}
\def \eea  {\end{eqnarray}}
\newcommand{\nn}{\nonumber}

\allowdisplaybreaks
\maketitle
\flushbottom

%%%%%%%%%%%%%%%%%%%%%%%%%%%%%%%%%%%%%%
\section{Introduction}
Since the advent of AGT correspondence \cite{Alday:2009aq} in 2009 by Alday, Gaiotto and Tachikawa,  the attempts on its proof never stop.  AGT correspondence reveals a connection  between 4D $\mathcal{N}=2$ gauge theory and 2D conformal field theory,  especially the instanton part of the Nekrasov partition function \cite{Nekrasov:2003af, Nekrasov:2003rj} is shown to be the same as the conformal blocks in correlation functions. In the original paper this is supported by instanton expansion up to certain nontrivial order, while its rigorous proof requires further efforts.  The $\Omega$ background introduced by Nekrasov has two deformation parameters $\e_1$ and $\e_2$, which simplifies when $\beta \equiv -\e_1/ \e_2=1$.
While some special cases like pure Yang-Mills case or other asymptotically free theories are already proved,  for the general conformal case with full flavors,  because of the mathematical difficulty,  only  $A_{1}$ group case at $\beta=1$ is done.  In this paper,  we focus on the proof of  its extension,  $A_{n}$ case at $\beta=1$.
AGT correspondence involves a wide range of research objects, and its generalized forms include: 5D N=1 gauge field and  2D q-deformed Toda field theory correspondence \cite{Awata:2009ur}, the Argyres-Douglas theory \cite{Argyres:1995jj} corresponding to the irregular conformal field theory, and the Nekrasov-Shatashvili limit (classical limit) \cite{Nekrasov:2009rc} and so on. 

There are two mainstream ways to prove the AGT correspondence. 
One approach is based on recursive relations.  For the conformal case,  Alba et al. \cite{Alba:2010qc} proposed that there should be a better basis for calculating the conformal block in the Verma additive group. Since the equivalence between the Nekrasov function and the conformal block must consider the contribution of the U(1) factor, the expanded Verma additive group introduces the Heisenberg algebra attached to the free field on the basis of the Virasoro/W algebra, and  the basis constructed by using their tensor product is called AFLT basis. In \cite{Alba:2010qc}, it is pointed out that the basis can also be expressed by Jack polynomials, and their inner product can be directly converted into the form of Nekrasov function in simple cases, and can also be proved by recursive relations in general cases. The main obstacle with this method is how to correctly interpret the AFLT basis.

Besides,  for pure gauge theories case,  in 2012 Schiffmann and Vasserot \cite{schiffmann2013cherednik} introduced the SH$^c$ algebra (spherical extension of degenerate double affine Hecke algebra), as a deformed version of the $W_{1+\infty}$ algebra. Using SH$^c$ algebra, they proved the AGT correspondence for pure Yang-Mills theory. An important advantage of the SH$^c$ algebra is that it performs well on the aforementioned AFLT basis, and can greatly simplify the complicated W algebra calculations. In fact, this property is also used as the recursive proof of AGT. The average value of the vertex operator under the AFLT base corresponds to the constituent elements of the Nekrasov instanton partition function, and after inserting the W algebra operator translated into the SH$^c$ operator, an infinity set of the recursive formula of the distribution function can be obtained, which is consistent with the conformal Ward identity of the conformal block of Toda field theory. This $A_{n}$ constraint verifies the AGT correspondence for a general quiver gauge theory \cite{Kanno:2013aha,  Bourgine:2015szm}.

The other approach is known as the direct proof,  which works well even for conformal case.  
In \cite{Mironov:2010pi} Mironov et al. managed to prove the $\beta=1$ case of $A_{1}$ AGT correspondence, by first transforming conformal blocks to Dostenko-Fatteev integral, rewriting as Selberg integral of two Schur polynomials, and finally showing its equivalence to the instanton part of Nekrasov partition function. However, for the case of $A_{n}$ with general $\beta$, what needs to be calculated is the integral of the product of $n+1$ Jack polynomials, which has no known formula in mathematics.  Furthermore,  the correspondence is no longer item-by-item,  but needs to be summed to eliminate redundant singularities, which already arise from $A_{1}$ with general $\beta$. In this regard, Morozov et al. \cite{Morozov:2013rma, Mironov:2013oaa} introduced the generalized Jack polynomial to avoid this issue,  yet the Selberg integral of this kind of polynomial still needs to be strictly derived.

In \cite{Zhang:2011au} Matsuo and one of the authors generalize this method to $A_{n}$ case with $\beta=1$, by conjecturing the Selberg integral of $n+1$ Schur polynomials, which is checked and confirmed by consistency relations. Then in 2020,  Albion et al. successfully derived the formula of this integral mathematically \cite{Albion:2020qhl}, but they regarded the integral conjecture in \cite{Zhang:2011au}  to be incorrect.  In this paper, we show that this formula is indeed the same as the conjecture,   after some equivalent transformation and taking care of the integration contour,  thus can be applied to the proof of AGT correspondence,  and now the $\beta=1$ $A_{n}$ case is rigorously proved.  To perform a double check, we also directly start from the formula in \cite{Albion:2020qhl}, and manage to prove the identification between the two sides of
AGT correspondence without using the conjecture made in  \cite{Zhang:2011au}.

In addition,  there is also an understanding of the AGT correspondence from the perspective of  geometric engineering. This explanation was proposed by Dijkgraaf and Vafa \cite{Dijkgraaf:2009pc}. Using the large N symmetry of topological strings, the spectral curve of the matrix model can be transformed into a Seiberg-Witten curve. Also, in the asymptotic free limit, the AGT correspondence in the case of less flavor can be proved by using the Zamolodchikov recursion formula. 
There are many other important results \cite{Bonelli:2010gk, Arnaudo:2022ivo} and interesting derivations from AGT correspondence. Such as the Gaiotto state \cite{Gaiotto:2009ma} and the qq-characters \cite{Nekrasov:2013xda}. There is also a reinterpretation of the AGT correspondence from a six-dimensional perspective \cite{Mironov:2015thk, Bonelli:2009zp}.

This paper is organized as follows.  In section 2 we provide the necessary information for AGT correspondence,  and illustrate the idea of the direct proof.  Section 3 briefly introduces Selberg integral,  and section 4 is the main part of this paper,  offering two equivalent approaches for the direct proof at $\beta=1$.  We conclude in section 5, and collect other details in the appendix.

\section{Roadmap for the direct proof of AGT correspondence}
The full Nekrasov partition function $Z_{\text{full}}$ of 4 dimensional $\mathcal{N}=2$ gauge theory has three components,
\ba
 Z_{\text{full}}= Z_{\text{tree}}
Z_{\text{1-loop}}Z_{\text{inst}},
\ea
with $Z_{\text{tree}}$ the classical (tree) part, $Z_{\text{1-loop}}$ the $1$-loop part and $Z_{\text{inst}}$ the instanton part.

On the other hand, the correlation function of 2 dimensional CFT reads,
\ba
{\text{Correlation function}}\sim {\text{Three-point functions}}
\times {\text{Conformal blocks}},
\ea
where the three-point functions can be computed by the DOZZ formula \cite{Dorn:1994xn,  Zamolodchikov:1995aa},  and conformal blocks can be obtained using the sewing procedure \cite{Belavin:1984vu, Sonoda:1988mf}.

AGT correspondence claims the identification between (the integration of) $Z_{\text{full}}$ and the correlation function. Especially, the relation between $Z_{\text{inst}}$ and the conformal blocks is checked by instanton expansion up to order 11 in the original paper. However, a strict proof is still needed, and the most direct way is to show analytically the equivalence between $Z_{\text{inst}}$ and the conformal blocks. Before doing that, let us dwell on some details on both sides.

\begin{figure}[H]
    \centering
    \includegraphics[width=0.48\textwidth]{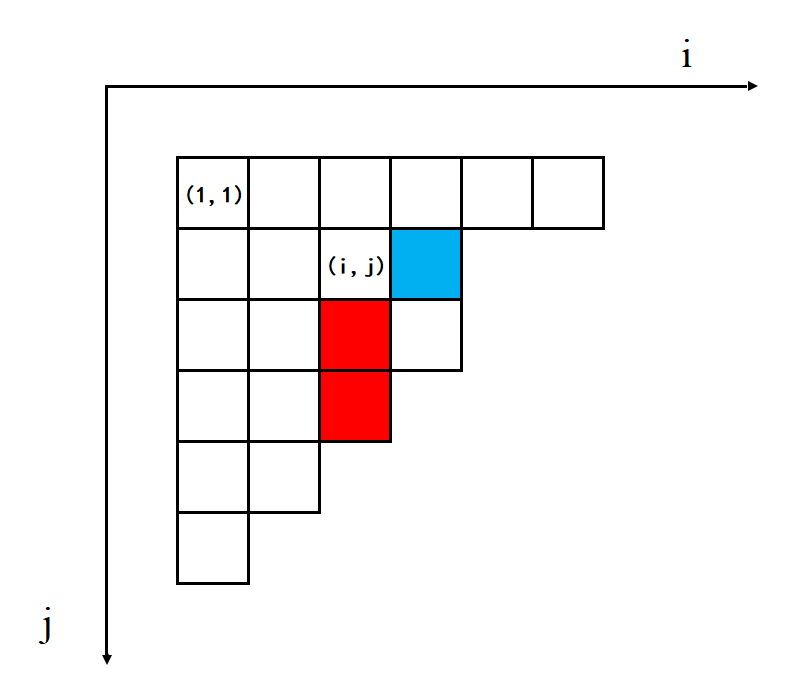}
    \caption{An example of Young diagram. For a box at the position $(i,j)$, Red boxes show its leg and Blue box(es) denote its arm.}
    \label{Young}
\end{figure}

For a $U(N)$ group, instantons can be labelled by a set of Young diagrams $\vec{Y}=\{Y^{(1)},Y^{(2)}, \dots , Y^{(N)}\}$.  As illustrated in Fig.\ref{Young},  a Young diagram stands for a collection of non-decreasing numbers $Y=(Y_1\geqslant Y_2\geqslant Y_3\geqslant \dots)$,  and in this sense $Y_i$ represents the height of the $i^\mathrm{th}$ column of $Y$,  while $Y'_i$ is the length of the $i^\mathrm{th}$ row. The leg length is defined by $Y_i-j$, and the arm length by $Y_j'-i$. The total length of $Y$ is $\ell_Y=\ell(Y)=Y_1^{'}$, and the total height is $h_Y=Y_1$. Later, we will use $L_Y$ to represent an arbitrary integer such that $L_Y \geqslant \ell_Y$.  The total number of boxes is named $| \vec Y|$. Each box $\square \in\vec{Y}$ can be distinguished by coordinates $x\equiv(r,i,j)$, so that it represents the (i,j) position of $Y^{(r)}$. We can also attach a parameter $\phi_x$ to it, defined as
\begin{equation}\label{phidef}
\phi_x=a_r+\b(i-1)-(j-1),
\end{equation}
where $a_r$ is the vacuum expectation value of the adjoint scalars in vector multiplets.

For a linear quiver with $\cM$ $A_{n-1}$ groups each associated with $\vec{Y}^I$, $1\leqslant I \leqslant \cM$, the instanton partition function reads:
\bea\label{Zinst}
Z_\text{inst}= &&\sum_{ \vec{Y}^1,\cdots \vec{Y}^{\cM}}\left(\prod_{I=1}^{\cM} q_I^{|  \vec Y^I|} \Zv(\vec a_I,\vec{Y}^I)\right)\prod_{i=1}^{n} \Zf(\vec a_1,\vec{Y}^1;  \m_i)\prod_{j=n+1}^{2n} Z_\text{afd}(\vec a_{\cM},\vec{Y}^{\cM};  \m_j) \nonumber \\
&&\times \prod_{I=1}^{\cM-1} \Zbf(\vec a_I,\vec{Y}^I;\vec a_{I+1}, \vec{Y}^{I+1}|m_{I,I+1}),
\eea
where $q_I$s are the gauge coupling constants, $ \vec Y^I$ indicates the fixed instanton, so that $q_I^{| \vec Y^I |} $ represents the instanton factor. $\vec a_I$ is the adjoint scalar vacuum expectation value (whose component is the one appeared in \eqref{phidef}),  $ \m_i$ is the mass of (anti-)fundamental hypermultiplet and $m_{I,I+1}$ stands for the mass of bifundamental hypermultiplet.

In this paper, we focus on the simple but  non-trivial case for a single $A_{n}$ with $2n+2$ flavors,
{\small
\bea\label{Ninst}
& Z_{\text{inst}}(q)=\sum_{\vec Y}
q^{|\vec Y|} N^{\text{inst}}_{\vec Y}( \vec a,  \mu),\\
& N^{\text{inst}}_{\vec Y}(  \vec a,  \mu)=z_{\mathrm{vect}}(\vec a,\vec Y)
\prod_{i=1}^{n+1} z_{\mathrm{fund}}(\vec a, \vec Y, \mu_i)
\prod_{j=n+2}^{2n+2}z_{\mathrm{afd}}(\vec a, \vec Y, \mu_j)=
 \frac{\prod_{r=1}^{n+1} \prod_{k=1}^{2n+2} f_{Y^{(r)}}(a_r+\mu_k )}
{\prod_{r,s=1}^{n+1} g_{Y^{(r)},W^{(s)}}(a_r - a_s)}, \nn
\eea
}
with $f_{Y}$ and $g_{Y,W}$ defined in the appendix.  Before the last equivalence we have actually redefined $-\m_i+\b-1$ in the fundamental parts as $\m_i$ so that they look the same as the anti-fundamental parts, for later convenience. 

On the other hand, for the CFT side,  the corresponding 4-point conformal block,  with insertion of screening charges   can be written as the Dotsenko-Fateev integral \cite{Dotsenko:1984nm,Itoyama:2010ki}
\be\label{4point}
\begin{split}
Z_\text{DF}(q) = 
&\left\langle \!\!\!\! \left\langle
: e^{( \tilde{\alpha_1}, \phi(0) )} :: e^{( \tilde{\alpha_2}, \phi(q) )} :
: e^{( \tilde{\alpha_3}, \phi(1) )} :: e^{( \tilde{\alpha_4}, \phi(\infty) )} :\right.\right. \\
&\left.\left.\times\prod_{a=1}^{n} \left(\int_0^q  : e^{b (e_a , \phi (z))}:d z \right)^{N_a}
\left(\int_1^{\infty}  : e^{b (e_a , \phi (z))}:d z \right)^{\tilde{N_a}}
\right\rangle\!\!\!\right\rangle \; ,
\end{split}
\ee
where $\langle \!\!\!\langle \quad \rangle\!\!\!\rangle $ is the matrix model average \cite{Itoyama:2009sc}.
Notice that there are some ambiguities about the integral contours of the screening charges, and here they are chosen as $0 \to q$ and $1\to \infty$.  However,  the correct contour choice in our situation is a little more complicated, as given in the next section.

After applying Wick's theorem,  
with $\tilde{\alpha_i} = \alpha_i / b $, $\beta =b^2 $ ,
(\ref{4point}) becomes \\
\begin{equation}
\begin{split}
& Z_\text{DF}(q) =q^{(\alpha_1 , \alpha_2)/ \beta} (1-q)^{(\alpha_2 , \alpha_3)/ \beta}
\prod_{a=1}^{n} \prod_{I=1}^{N_a}\int_0^q d z_I^{(a)}\prod_{J=N_a +1}^{N_a + \tilde{N_a}}\int_1^{\infty}d z_J^{(a)}
\prod_{i<j}^{N_a + \tilde{N_a}} (z_j^{(a)} - z_i^{(a)})^{2 \beta} \quad \times
\\ &\times
\prod_i^{N_a + \tilde{N_a}} (z_i^{(a)} )^{(\alpha_1 , e_a)}
(z_i^{(a)} - q )^{(\alpha_2 , e_a)} (z_i^{(a)}-1 )^{(\alpha_3 , e_a)}
%\quad \times
%\\ &\times
\prod_{a=1}^{n-1}\prod_i^{N_a + \tilde{N_{a}}} \prod_j^{N_{a+1} + {\tilde N}_{a+1}}(z_j^{(a+1)} - z_i^{(a)})^{- \beta} \quad .
\end{split}\label{ZDF}
\end{equation}
This equation can be further written in terms of Selberg integral introduced in the next section (after getting rid of the $U(1)$ factor,  and taking care of the contour) \cite{Zhang:2011au},
\begin{equation}
\tilde Z_\text{DF}(q)=
\sum_{\vec{Y}}
 q^{\lvert \vec{Y} \rvert}
\left\langle
\prod_{a=1}^{n}
j^{(\beta)}_{Y_a} (p_k^{(a-1)}-p_k^{(a)} - \frac{v'_{a+}}{\beta})
\right\rangle_{+}
\left\langle
\prod_{a=1}^{n}
j^{(\beta)}_{Y_a} (\tilde{p}_k^{(a)}-\tilde{p}_k^{(a-1)}+ \frac{v'_{a-}}{\beta})
\right\rangle_{-} \; ,
\label{DF-S}
\end{equation}
where $\langle \rangle_{\pm}$ means taking Selberg integral with respect to Selberg average with certain set of parameters denoted by $\pm$ for short,  as given by \eqref{opm}.

To prove AGT correspondence,  we just need to show that 
\begin{equation}
\label{instCB}
Z_{\text{inst}}(q)=\tilde Z_\text{DF}(q), 
\end{equation}
with Selberg integral serving as an intermediate step.

\section{Selberg integral}
This section mainly summarizes the results in \cite{Albion:2020qhl}.
The Selberg integral \cite{forrester2008importance} is defined as
\begin{align}\label{Eq_Selberg}
S_N(u,v;\b)&\equiv
\Int_{[0,1]^N}\! \prod_{i=1}^N x_i^{u}(1-x_i)^{v}
\prod_{1\leqslant i<j\leqslant N}
|{x_i-x_j}|^{2\b}
\,\mathrm{d}x_1\cdots \mathrm{d} x_N \\
&\hphantom{:}=\prod_{i=1}^N
\frac{\Gamma(v+1+(i-1)\b)\Gamma(u+1+(i-1)\b)\Gamma(1+i\b)}
{\Gamma(u+v+2+(2N-i-1)\b)\Gamma(1+\b)}, \notag 
\end{align}
with $\mathrm{Re}(u+1)>0$, $\mathrm{Re}(v+1)>0$ and
$
\mathrm{Re}(\b)>-\min\{1/N,\mathrm{Re}(u+1)/(N-1),
\mathrm{Re}(v+1)/(N-1)\}.
$

Its $A_n$ extension, with an insertion of a polynomial
$\mathscr{O}(x^{(1)},\dots,  x^{(n)})$ symmetric in each of the alphabets
$x^{(r)}$,  is given by
\begin{align}\label{Eq_I-An}
I^{A_n}_{N_1,\dots,N_n}&(\mathcal{O};u_1,\dots,u_n,v;\b) \\
&\equiv\Int_{C_{\b}^{N_1,\dots,N_n}[0,1]}
\mathcal{O}\big(x^{(1)},\dots,x^{(n)}\big)
\prod_{r=1}^n \prod_{i=1}^{N_r}\big(x^{(r)}_i\big)^{u_r}
\big(1-x^{(r)}_i\big)^{v_r} \notag \\[-1mm]
&\qquad\qquad\qquad\quad\times
\prod_{r=1}^n \Abs{\Delta\big(x^{(r)}\big)}^{2\b}
\prod_{r=1}^{n-1} \Abs{\Delta\big(x^{(r)},x^{(r+1)}\big)}^{-\b}\, 
\dup x^{(1)}\cdots\dup x^{(n)}, \notag
\end{align}
where
the contour $C_{\b}^{N_1,\dots,N_n}[0,1]$ is a real domain chain,  and 
\begin{equation}\label{Eq_beta-r}
v_1=\dots=v_{n-1}:=0, \quad v_n:=v,
\end{equation}
with
\begin{gather}\label{Eq_conditions-a}
\abs{\Re(\b)}<\frac{1}{N_n},\quad
\Re(v+1)>0, \quad
\Re\big(v+1+(N_n-1)\b\big)>0, \\[2mm]
\Re\big(u_r+\dots+u_s+(r-s-1)(\b-1)+i\b\big)>0
\quad\text{for $1\leqslant r\leqslant s\leqslant n$ and $1\leqslant i\leqslant N_r-N_{r-1}$}.
\end{gather}
For $x,y$ two alphabets which respectively have length of $N_1$ and $N_2$, the Vandermonde products are defined as $\Delta(x)$ and $\Delta(x,y)$,
\be\label{Eq_Vandermonde}
\D(x)\equiv\prod_{1\leqslant i<j\leqslant N_1}(x_i-x_j)\quad
\text{and}\quad
\D(x,y)\equiv\prod_{i=1}^{N_1}\prod_{j=1}^{N_2}(x_i-y_j).
\ee

Then the $A_n$ Selberg average of a symmetric polynomial 
$\mathscr{O}$ is defined as
\begin{equation}\label{Eq_An-Selberg-average}
\big\langle \mathcal{O}\big\rangle_{u_1,\dots,u_n,v;\b}
^{N_1,\dots,N_n}\equiv\frac{I^{A_n}_{N_1,\dots,N_n}
(\mathcal{O};u_1,\dots,u_n,v;\b)}
{I^{A_n}_{N_1,\dots,N_n}(1;u_1,\dots,u_n,v;\b)}.
\end{equation} 
When one want to consider the $\b \to 1$ limit of this integral,  noticing  \eqref{Eq_conditions-a} implies that 
$\lvert \Re(\b) \rvert <1$,
the once real domain
$C_{\b}^{N_1,\dots,N_n}[0,1]$ in this integral must be replaced by the complex contour defined as follows
\begin{equation}\label{Eq_contour}
C^{N_1,\dots,N_n}=C_1^{N_1}\times\dots\times C_n^{N_n},
\quad \text{where}\quad 
C_r^{N_r}=\underbrace{C_r\times\dots\times C_r}_{N_r \textrm{ times}}.
\end{equation}

Each positively oriented curve $C_r$ passes through the origin and contains the interval $(0,1]$ in its interior. The contour $C_r$ is contained in the interior of $C_{r-1}$ for $r\geqslant 2$. Later we will use this complex contour $C^{N_1,\dots,N_n}$ to replace the domain  of integration in Selberg integral form of the Dotsenko-Fateev integral \eqref{4point}. 

For $\b=1$, the special case concerned in this paper ,  choosing $0\leqslant N_1\leqslant \cdots\leqslant N_n$ and 
$u_1,\dots,u_n,v\in\mathbb{C}$ such that
\begin{equation}\label{Eq_conv}
\Re(u_r+\cdots+u_s+1)>0 \quad
\text{for $1\leqslant r\leqslant s\leqslant n$},
\end{equation}
we now define the $\b=1$ $A_n$ Selberg average as follows:
\bea
\big\langle \mathscr{O}\big\rangle_{u_1,\dots,u_n,v}
^{N_1,\dots,N_n}\equiv
\frac{I^{A_n}_{N_1,\dots,N_n}
(\mathscr{O};u_1,\dots,u_n,v)}
{I^{A_n}_{N_1,\dots,N_n}(1;u_1,\dots,u_n,v)},
\eea
and 
\bea\label{opm}
\big\langle \mathscr{O}\big\rangle_\pm\equiv\big\langle \mathscr{O}\big\rangle_{u_{1\pm},\dots,u_{n\pm},v_\pm}
^{N_1,\dots,N_n},
\eea
where, assuming \eqref{Eq_beta-r}, 
\begin{align}\label{Eq_I-def}
&I^{A_n}_{N_1,\dots,N_n}(\mathscr{O};u_1,\dots,u_n,v) \\
&\qquad \equiv\frac{1}{(2\pi\iup)^{N_1+\dots+N_n}} \Int_{C^{N_1,\dots,N_n}}
\mathscr{O}\big(\xar{1},\dots,\xar{n}\big)
\prod_{r=1}^n \prod_{i=1}^{N_r}\big(\xar{r}_i\big)^{u_r}
\big(\xar{r}_i-1\big)^{v_r} \notag \\[-1mm]
&\qquad\qquad\qquad\qquad\qquad\qquad\qquad\times
\prod_{r=1}^n \Delta^2\big(\xar{r}\big)
\prod_{r=1}^{n-1} \Delta^{-1}\big(\xar{r},\xar{r+1}\big)\, 
\dup\xar{1}\cdots\dup\xar{n}. \notag
\end{align}

%\begin{theorem}\label{Thm_nplusone}
For $n$ a positive integer, let $0\leqslant N_1\leqslant\cdots\leqslant N_n$ be
integers, $u_1,\dots,u_n,v\in\mathbb{C}$ such that
\eqref{Eq_conv} holds, and $\Yar{1},\dots,\Yar{n+1}\in\mathscr{P}$.
Besides, set $\xar{0}\equiv0$ and $\xar{n+1}\equiv-v$.
Then,  the Selberg integral average of $n+1$ Schur polynomials $\chi_Y$ is given by the following theorem:  
\paragraph{Theorem}
{\em In \cite{Albion:2020qhl}  it is proved that
} 
\begin{align}
\label{Eq_nplusone}
&\bigg\langle 
\prod_{r=1}^{n+1} \chi_{\Yar{r}}\big[\xar{r}-\xar{r-1}\big]
\bigg\rangle_{u_1,\dots,u_n,v}^{N_1,\dots,N_n} \\
&\qquad=
\prod_{r=1}^{n+1}\prod_{1\leqslant i<j\leqslant L_{\Yar{r}}} 
\frac{\Yar{r}_i-\Yar{r}_j+j-i}{j-i} 
\prod_{r,s=1}^{n+1}\prod_{i=1}^{L_{\Yar{r}}}
\frac{(A_{r,s}-N_{s-1}+N_s-i+1)_{\Yar{r}_i}}
{(A_{r,s}+L_{\Yar{s}}-i+1)_{\Yar{r}_i}} \notag \\
&\qquad\quad\times 
\prod_{1\leqslant r<s\leqslant n+1}\prod_{i=1}^{L_{\Yar{r}}}\prod_{j=1}^{L_{\Yar{s}}}
\frac{\Yar{r}_i-\Yar{s}_j+A_{r,s}+j-i}{A_{r,s}+j-i}, \notag
\end{align}
with notations collected in appendix \ref{can}.\par 
This theorem will be applied in the direct proof in the next section.

\section{Proof for $\beta=1$, $A_{n}$ case}
 We finalize the proof for $A_{n}$ AGT correspondence at $\beta=1$ in this section,  with the help of the strict theorem \eqref{Eq_nplusone} for the $A_{n}$ Selberg integral containing a product of $n+1$ Schur polynomials.

Considering the explicit expression of $Z_{\text{inst}}$ \eqref{Ninst}, to prove $A_{n}$ AGT correspondence at $\b=1$, we just need to confirm that
\begin{equation}\label{agt-1}
\begin{split}
    &\left<\prod_{r=1}^{n+1}\chi_{Y^{(r)}}\left[x^{(r-1)}-x^{(r)}-v^{'}_{r+}\right]\right>_+\left<\prod_{r=1}^{n+1}\chi_{Y^{(r)}}\left[y^{(r)}-y^{(r-1)}+v^{'}_{r-}\right]\right>_-\\=&\left.\frac{\prod_{r=1}^{2n+2}\prod_{s=1}^{n+1}f_{Y^{(s)}}(\mu_r+a_s)}{\prod_{r,s=1}^{n+1}G_{Y^{(r)},Y^{(s)}}(a_r-a_s)G_{Y^{(s)},Y^{(r)}}(a_s-a_r+1-\b)}\right|_{\beta=1}.
\end{split}
\end{equation}
Here we  explain some notations:
\begin{align*}
    v^{'}_{(n+1-r)+}\equiv\sum_{s=1}^r{v_{(n+1-s)+}}, \qquad v^{'}_{r-}\equiv-\sum_{s=1}^{r-1}{v_{s-}};
\end{align*}
\begin{align*}
    x^{(0)}=x^{(n+1)}=y^{(0)}=y^{(n+1)}=N_{0+}=N_{(n+1)+}=N_{0-}=N_{(n+1)-}=0.
\end{align*}

Now we have two ways to prove \eqref{agt-1}. 
\paragraph{Method 1:}

The first one is to verify the validity of the conjecture proposed in \cite{Zhang:2011au}, as shown below.

\paragraph{Conjecture }
{\em In \cite{Zhang:2011au}  the following formula of Selberg average for $n+1$ Schur
polynomials is proposed,
} 
\bea
\label{nschur}
&\left\langle
\chi_{Y_1} (-p_k^{(1)} - v_1')\dots
\chi_{Y_r} (p_k^{(r-1)} -p_k^{(r)} - v_r')\dots
\chi_{Y_{n+1}} (p_k^{(n)})
 \right\rangle^{A_{n}}_{\vec u,\vec v,\beta=1}\nn
\\
&=\prod_{s=1}^{n}
\bigg\{
(-1)^{|Y_s|}\; \times
\frac{[v_{s}+ N_s - N_{s-1} ]_{Y'_s} }
{[ N_s + N_{s-1}  ]_{Y'_s} }
 \; \times \!
\prod_{1\leqslant i<j\leqslant N_{s-1} + N_s}
\frac{(j-i+1)_{Y'_{si}-Y'_{sj}}}{(j-i)_{Y'_{si}-Y'_{sj}}}
\bigg\}
\nn\\
&\times
\prod_{1\leqslant i<j\leqslant N_{n}}
\frac{(j-i+1)_{Y_{(n+1)i}-Y_{(n+1)j}}}{(j-i)_{Y_{(n+1)i}-Y_{(n+1)j}}}
\\
&\times
\prod_{1\leqslant t<s\leqslant n+1}
\bigg\{
\frac{[ v_{t}+u_{t}+\dots +u_{s-1} + N_t- N_{t-1}   ]_{Y'_t} }
{[ v_{t}- v_{s}+u_{t}+\dots +u_{s-1}+ N_{t} - N_{t-1} - N_s  ]_{Y'_t} }
\nn\\
&\times
\frac{[-v_{s}+u_{t}+\dots +u_{s-1} - N_s + N_{s-1} ]_{Y_s} }
{[ v_{t}- v_{s}+u_{t}+\dots +u_{s-1} - N_{t-1} - N_s + N_{s-1}  ]_{Y_s} }
 \nn\\
&\qquad\times
\prod_{i=1}^{N_{t}}\prod_{j=1}^{N_{s-1}}
\frac{  v_{t}- v_{s}+u_{t}+\dots +u_{s-1} + N_t- N_{t-1} - N_s + N_{s-1}  +1 -(i +j)}
{ v_{t}- v_{s}+u_{t}+\dots +u_{s-1} + N_t- N_{t-1} - N_s + N_{s-1}   +1 + Y'_{ti} + Y_{sj} -(i +j) }
\bigg\} \;,\nn
\eea
with $v_r'\equiv\sum_{a=r}^n v_a=v\delta_{r1}$ after imposing the constraint $v_2=\cdots=v_{n}=0$, and $v_1=v$. For $x$ a number and $Y$ a Young diagram, $[x]_Y$ is defined as,
\[
[x]_Y=\prod_{(i,j)\in Y}(x-i+j).
\]

This conjecture's equivalence to the theorem \eqref{Eq_nplusone} is justified in appendix \ref{coc}.
Since in \cite{Zhang:2011au}, it is already confirmed that the validity of this conjecture will result in the proof of AGT correspondence for $\b =1$,  the issue is settled.

\paragraph{Method 2:}

Another way is to prove \eqref{agt-1} directly by \eqref{Eq_nplusone}, which is given as follows.
Firstly we adopt the restriction on parameters $v$,  as required by Toda field theory \cite{Fateev:2007ab,  Fateev:2008bm},
\begin{gather}
    v_{r+}=v_+\delta_{r1}\text{ and }v_{r-}=v_-\delta_{rn}.
\end{gather}

For convenience,  we redefine $N_{0+}=x^{(0)}=-v_+$ and $N_{(n+1)-}=y^{(n+1)}=-v_-$. 
Then we can rewrite \eqref{agt-1} as
\begin{equation}\label{proof-step1}
\begin{split}
    &\left<\prod_{r=1}^{n+1}\chi_{Y^{(r)}}\left[x^{(r-1)}-x^{(r)}\right]\right>_+\left<\prod_{r=1}^{n+1}\chi_{Y^{(r)}}\left[y^{(r)}-y^{(r-1)}\right]\right>_-\\
    =&\left.\frac{\prod_{r=1}^{2n+2}\prod_{s=1}^{n+1}f_{Y^{(s)}}(\mu_r+a_s)}{\prod_{r,s=1}^{n+1}G_{Y^{(r)},Y^{(s)}}(a_r-a_s)G_{Y^{(s)},Y^{(r)}}(a_s-a_r+1-\b)}\right|_{\beta=1}.
\end{split}
\end{equation}
The sufficient condition can be chosen as that the left two Selberg averages are respectively equal to two parts of the right hand side in \eqref{proof-step1}. Specifically,
\begin{equation}\label{x-part}
    \left<\prod_{r=1}^{n+1}\chi_{Y^{(r)}}\left[x^{(r-1)}-x^{(r)}\right]\right>_+=\left.\frac{\prod_{s=1}^{n+1}(-1)^{|{Y^{(s)}}|}\prod_{r=1}^{n+1}\prod_{s=1}^{n+1}f_{Y^{(s)}}(\mu_r+a_s)}{\prod_{r,s=1}^{n+1}G_{Y^{(r)},Y^{(s)}}(a_r-a_s)}\right|_{\beta=1},
\end{equation}
and the second part,
\begin{equation}\label{y-part}
    \left<\prod_{r=1}^{n+1}\chi_{Y^{(r)}}\left[y^{(r)}-y^{(r-1)}\right]\right>_-=\left.\frac{\prod_{s=1}^{n+1}(-1)^{|{Y^{(s)}}|}\prod_{r=n+2}^{2n+2}\prod_{s=1}^{n+1}f_{Y^{(s)}}(\mu_r+a_s)}{\prod_{r,s=1}^{n+1}G_{Y^{(r)},Y^{(s)}}(a_r-a_s)}\right|_{\beta=1}.
\end{equation}
Notice $r$ and $s$ are dummy indexes, we exchange them in \eqref{y-part}. These two equations are the main claim of the paper.   Their proofs are alike,  and here we first check \eqref{y-part}.  This can be realized by deforming the Theorem \eqref{Eq_nplusone},  and applying several lemmas introduced below.

We decompose the right hand side of \eqref{Eq_nplusone} into three parts for further calculation:
\begin{equation}\label{key-step}
\begin{split}
    &\prod_{r=1}^{n+1}\prod_{1\leqslant i<j\leqslant L_{\Yar{r}}}\frac{Y^{(r)}_i-Y^{(r)}_j+j-i}{j-i}\prod_{r,s=1}^{n+1}\prod_{i=1}^{L_{\Yar{r}}}\frac{(A_{r,s}-N_{s-1}+N_{s}-i+1)_{Y^{(r)}_i}}{(A_{r,s}+L_{\Yar{s}}-i+1)_{Y^{(r)}_i}}\\
    \times&\prod_{1\leqslant r<s\leqslant n+1}\prod_{i=1}^{L_{\Yar{r}}}\prod_{j=1}^{L_{\Yar{s}}}\frac{Y^{(r)}_i-Y^{(s)}_j+A_{r,s}+j-i}{A_{r,s}+j-i}\\
    =&\prod_{r=1}^{n+1}\left(\prod_{1\leqslant i<j\leqslant L_{\Yar{r}}}\frac{Y^{(r)}_i-Y^{(r)}_j+j-i}{j-i}\prod_{i=1}^{L_{\Yar{r}}}\frac{1}{(L_{\Yar{r}}-i+1)_{Y^{(r)}_i}}\right)\\
    \times&\prod_{r,s=1}^{n+1}\prod_{i=1}^{L_{\Yar{r}}}(A_{r,s}-N_{s-1}+N_{s}-i+1)_{Y^{(r)}_i}\\
    \times&\prod_{1\leqslant r<s\leqslant n+1}\left(\prod_{i=1}^{L_{\Yar{r}}}\right.\prod_{j=1}^{L_{\Yar{s}}}\frac{Y^{(r)}_i-Y^{(s)}_j+A_{r,s}+j-i}{A_{r,s}+j-i}\\&\prod_{i=1}^{L_{\Yar{r}}}\frac{1}{(A_{r,s}+L_{\Yar{s}}-i+1)_{Y^{(r)}_i}}\left.\prod_{i=1}^{L_{\Yar{s}}}\frac{1}{(A_{s,r}+L_{\Yar{r}}-i+1)_{Y^{(s)}_i}}\right).
\end{split}
\end{equation}
The main step is the decomposition of $\prod_{r,s=1}^{n+1}\prod_{i=1}^{L_{\Yar{r}}}\frac{1}{(A_{r,s}+L_{\Yar{s}}-i+1)_{Y^{(r)}_i}}$ shown as follows
\bea
    &\prod_{r,s=1}^{n+1}\prod_{i=1}^{L_{\Yar{r}}}\frac{1}{(A_{r,s}+L_{\Yar{s}}-i+1)_{Y^{(r)}_i}}\nn\\
    =&\prod_{r(=s)=1}^{n+1}\prod_{i=1}^{L_{\Yar{r}}}\frac{1}{(A_{r,r}+L_{\Yar{r}}-i+1)_{Y^{(r)}_i}}\prod_{1\leqslant r<s\leqslant n+1}\prod_{1\leqslant s<r\leqslant n+1}\prod_{i=1}^{L_{\Yar{r}}}\frac{1}{(A_{r,s}+L_{\Yar{s}}-i+1)_{Y^{(r)}_i}}\\
    =&\prod_{r=1}^{n+1}\prod_{i=1}^{L_{\Yar{r}}}\frac{1}{(L_{\Yar{r}}-i+1)_{Y^{(r)}_i}}\prod_{1\leqslant r<s\leqslant n+1}\prod_{i=1}^{L_{\Yar{r}}}\frac{1}{(A_{r,s}+L_{\Yar{s}}-i+1)_{Y^{(r)}_i}}\prod_{i=1}^{L_{\Yar{s}}}\frac{1}{(A_{s,r}+L_{\Yar{r}}-i+1)_{Y^{(s)}_i}}.\nn
\eea
In the last step, we use $A_{r,r}=0$ and switch part of $r$ with $s$.
Equation \eqref{key-step} can be transformed to a form closer to the right hand side of \eqref{agt-1}, with the help of the following lemmas proved in the appendix.\\
\textbf{Lemma 1}
\begin{equation}
    \prod_{1\leqslant i<j\leqslant L_Y}\frac{(Y_i-Y_j+\b(j-i))_{\b}}{(\b(j-i))_{\b}}\prod_{i=1}^{L_Y}\frac{1}{(\b(L_Y-i+1))_{Y_i}}
    =\frac{1}{G_{Y,Y}(0)}.
\end{equation}
\textbf{Lemma 2}
\begin{equation}
    \prod_{i = 1}^{L_Y}(-z-\b i+\b)_{Y_i}=(-1)^{|Y|}f_{Y}(z).
\end{equation}
\textbf{Lemma 3}
\begin{equation}
\begin{split}
    \prod_{i=1}^{L_Y}\prod_{j=1}^{L_W}\frac{Y_i-W_j-x+j-i}{-x+j-i}
        &\prod_{i=1}^{L_Y}\frac{1}{(-x+L_W-i+1)_{Y_i}}
        \prod_{i=1}^{L_W}\frac{1}{(x+L_Y-i+1)_{W_i}}\\
        =&\left.\frac{(-1)^{|Y|+|W|}}{G_{Y,W}(x)G_{W,Y}(-x)}\right|_{\beta=1}.
    \end{split}
\end{equation}
The left hand sides of these three lemmas at $\b=1$ are respectively corresponding to the three terms in equation \eqref{key-step}. By applying the lemmas, we get
\be\label{general}
\begin{split}
&\left<\prod_{r=1}^{n+1}\chi_{W^{(r)}}\left[z^{(r)}-z^{(r-1)}\right]\right>_{u_1,...,u_n,v}^{N_1,...,N_n}\\
=&\prod_{r=1}^{n+1}\left(\frac{1}{G_{W^{(r)},W^{(r)}}(0)}\right)
\times \prod_{r,s=1}^{n+1}(-1)^{|W^{(r)}|}f_{W^{(r)}}(-(A_{r,s}-N_{s-1}+N_s))\\\times&\prod_{1\leqslant r<s\leqslant n+1}\frac{(-1)^{|W^{(r)}|+|W^{(s)}|}}{G_{W^{(r)},W^{(s)}}(A_{s,r})G_{W^{(s)},W^{(r)}}(A_{r,s})}\\
=&\prod_{r=1}^{n+1}(-1)^{|W^{(r)}|}\prod_{r,s=1}^{n+1}\frac{f_{W^{(r)}}(-(A_{r,s}-N_{s-1}+N_s))}{G_{W^{(r)},W^{(s)}}(A_{s,r})}.
\end{split}
\ee
To avoid confusion of notations, we use $W$ instead of $Y$ to represent the Young diagrams in \eqref{general} and $z$ instead of $x$. The exponent of $-1$ should be calculated carefully,
\be
\prod_{r,s=1}^{n+1}(-1)^{|W^{(r)}|}\prod_{1\leqslant r<s\leqslant n+1}(-1)^{|W^{(r)}|+|W^{(s)}|}=\prod_{r=1}^{n+1}(-1)^{(n+1)|W^{(r)}|}\times(-1)^{n|W^{(r)}|}=\prod_{r=1}^{n+1}(-1)^{|W^{(r)}|}.
\ee
Formula \eqref{general} can be applied to integral in \eqref{y-part} directly by setting notations agree. We take $z^{(r)}=y^{(r)}$, $N_r=N_{r-}$, $W^{(r)}=Y^{(r)}$, $u_r=u_{r-}$ for any $0\leqslant r\leqslant n+1$ and $v=v_-$.
\bea
         &\left<\prod_{r=1}^{n+1}\chi_{Y^{(r)}}\left[y^{(r)}-y^{(r-1)}\right]\right>_-
         =\prod_{r=1}^{n+1}(-1)^{|Y^{(r)}|}\times\\&\prod_{r,s=1}^{n+1}\frac{f_{Y^{(r)}}(-(\sum_{i=r}^{s-1}u_{i-}-\sum_{i=s}^{r-1}u_{i-}+N_{r-}-N_{(r-1)-}-N_{s-}+N_{(s-1)-}-N_{(s-1)-}+N_{s-}))}{G_{Y^{(r)},Y^{(s)}}(-(\sum_{i=r}^{s-1}u_{i-}-\sum_{i=s}^{r-1}u_{i-}+N_{r-}-N_{(r-1)-}-N_{s-}+N_{(s-1)-}))}.\nn
    \eea
To see the equivalence, we give following identifications of parameters
\begin{equation}
    \begin{split}
        \mu_{s+n+1}+a_r&=-\left(\sum_{i=r}^{s-1}u_{i-}-\sum_{i=s}^{r-1}u_{i-}+N_{r-}-N_{(r-1)-}\right);\\
        a_r-a_s&=-\left(\sum_{i=r}^{s-1}u_{i-}-\sum_{i=s}^{r-1}u_{i-}+N_{r-}-N_{(r-1)-}-N_{s-}+N_{(s-1)-}\right).
    \end{split}
\end{equation}
$r$ and $s$ run from $1$ to $n+1$.
Then we have
\begin{equation}
    \begin{split}
    \left<\prod_{r=1}^{n+1}\chi_{Y^{(r)}}\left[y^{(r)}-y^{(r-1)}\right]\right>_-&=\prod_{r=1}^{n+1}(-1)^{|Y^{(r)}|}\prod_{r,s=1}^{n+1}\frac{f_{Y^{(r)}}(\mu_{s+n+1}+a_r)}{G_{Y^{(r)},Y^{(s)}}(a_r-a_s)}\\
    &=\frac{\prod_{s=1}^{n+1}(-1)^{|{Y^{(s)}}|}\prod_{r=n+2}^{2n+2}\prod_{s=1}^{n+1}f_{Y^{(s)}}(\mu_r+a_s)}{\prod_{r,s=1}^{n+1}G_{Y^{(r)},Y^{(s)}}(a_r-a_s)},
    \end{split}
\end{equation}
which is indeed equation \eqref{y-part}.\par

The proof for \eqref{x-part} is parallel but a little more extented,  which is given in appendix \ref{app-x}.  With both equations verified, we find that \eqref{proof-step1} is valid,  and the $A_{n}$ AGT correspondence at $\beta=1$ is proved.
\section{Conclusion and future directions}
In this paper, we complete the direct proof for $A_n$ AGT correspondence in a particular case of $\b=1$,  following the method proposed in \cite{Mironov:2010pi} by Mironov et al.  with the help of the strict formula for $A_n$ Selberg integral of Schur polynomials proved in \cite{Albion:2020qhl}. We provide two approaches for this proof. One is to confirm the conjecture given in \cite{Zhang:2011au}, which can lead to AGT relation. The other is to transform the four-point function to Selberg integral and prove its equality to the Nekrasov partition function directly. 

This work can be generalized in several ways. First, the simplest nontrivial case, $G=A_n$, with $N_f=2n+2$ hypermultiplets in fundamental representation can be generalized to linear quiver gauge theory with gauge group $G=A_{n_1}\times\dots\times A_{n_k}$. Secondly,  for general $\b$, even $A_1$ AGT correspondence has not been strictly proved because of the necessity of extra poles cancellation.  The once term-wise relation of \eqref{instCB} for each Young diagram now needs to be summed over at general  $\b$.  Morozov et al.  \cite{Morozov:2013rma,  Mironov:2013oaa} defined the so-called generalized Jack polynomials to circumvent this issue.
However, the Selberg integral for generalized Jack polynomials is still open and we may discover some other methods to avoid the extra poles problem.  Besides,  since AGT correspondence has an extension to 5D Gauge theory and 2D q-CFT,  the proof of this case can also be addressed.

\bigskip
\bigskip

\noindent {\it Acknowledgements.} 
The authors thank S. Ole Warnaar, Yutaka Matsuo, Jean-Emile  Bourgine,  Rui-Dong Zhu,  Hong-Fei Shu, Yang Lei and Qian Shen for helpful suggestions.  KZ (Hong Zhang) thanks Seamus P. Albion for bring back the attention to this interesting subject.
This work is supported by a classified fund of Shanghai city.  
\appendix
\section{Notations and conventions}\label{can}
The components of \eqref{Zinst} are defined as the following:
First the bifundamental part,
\begin{eqnarray}\label{Zbfd}
\Zbf(\vec{a},\vec Y;\vec{b},\vec{W}|m_{1,2})
&=& \prod_{r=1}^{N_1} \prod_{s=1}^{N_2} g_{Y^{(r)},W^{(s)}}(a_r-b_{s}-m_{1,2})\,,\\
g_{Y,W}(z)&=&G_{Y,W}(z)G_{W,Y}(-z+1-\b)\\
G_{Y,W}(z)&=&\prod_{(i,j)\in Y}\left(z+\b(Y_j'-i)+(W_i-j)+\b\right)
\end{eqnarray}
And the vector part can be given as
\begin{eqnarray}
\Zv(\vec{a},\vec{Y})&\equiv& \Zbf(\vec{a},\vec{Y};\vec{a},\vec{Y}|0)^{-1}.
\end{eqnarray}
The fundamental part is
\begin{eqnarray}\label{def_Zf}
Z_\mathrm{fund}(\vec{a},\vec{Y};m)=\prod_{x\in \vec{Y}}(\phi_x-m+\b-1)&=&\prod_{r=1}^{N}\prod_{(i,j)\in Y^{(r)}}\Big(a_r-m+\b-1+\b(i-1)-(j-1)\Big)\nn\\
=\prod_{r=1}^{N} f_{Y^{(r)}}(a_r-m+\b-1),
\end{eqnarray}
and the anti-fundamental part is
\begin{eqnarray}\label{def_Zaf}
Z_\mathrm{afd}(\vec{a},\vec{Y};{m})=\prod_{x\in \vec{Y}}(\phi_x+m)&&=\prod_{r=1}^{N}\prod_{(i,j)\in Y^{(r)}}\Big(a_r+m+\b(i-1)-(j-1)\Big)\nn\\
=\prod_{r=1}^{N} f_{Y^{(r)}}(a_r+m)&&\equiv Z_\mathrm{afd}(\vec{a},\vec{Y};{-m+\b-1}),
\end{eqnarray}
with
\begin{align}
f_Y(z) = \prod\limits_{(i,j)\in Y}\Big(z+\b(i-1)-(j-1)\Big)\; .
\end{align}

For $x$ an indeterminate or complex number and $k$ a complex number, the Pochhammer symbol $(x)_k$ is defined as
\begin{equation}
    (x)_k\equiv\frac{\Gamma(x+k)}{\Gamma(x)}.
\end{equation}
There is a shorthand notation,
\begin{equation}
    A_{r,s}\equiv u_r+...+u_{s-1}+N_r-N_{r-1}-N_{s}+N_{s-1},
\end{equation}
for $1\leqslant r<s\leqslant n+1$, and $A_{s,r}=-A_{r,s}$ for any $r$ and $s$, in particular, $A_{r,r}=0$. $N_r$ is the cardinality of $\{x^{(r)}\}$ for $1\leqslant N\leqslant n$. $N_0=x^{(0)}=0$ and $N_{n+1}=x^{(n+1)}=-v$. The explicit form of $A_{r,s}$ for any $r$ and $s$ is given here
\bea
    A_{r,s}=\sum_{i=r}^{s-1}u_i-\sum_{i=s}^{r-1}u_i+N_r-N_{r-1}-N_s+N_{s-1},
\eea
Notice that $\sum_{i=r}^{s}f=0$ for any expression $f$ when $s<r$. For products, the result is 1. i.e.,
\begin{equation}\label{one}
    \prod_{i=r}^{s}f=1, \; when\, s<r.
\end{equation}
%\end{theorem}
\textbf{Jack polynomials and Schur Functions}\label{a:Jack}\par
Jack polynomials $J^{(\beta)}_Y[x_1,\cdots,x_M]$ 
form a family of
symmetric polynomials labeled by Young diagrams $Y$ and depending on the
variables
 $x_1,\cdots, x_M$.
The paper \cite{stanley1989some} provides some detailed properties of Jack polynomial. 
We use the power sum $p_k(z) = \sum_i x_i^k$ to denote the Jack polynomial, with
$J^{(\beta)}_Y(p_1, p_2,\cdots)\equiv J^{(\beta)}_Y[x_1,\cdots, x_M]$.
The explicit forms of the first two level Jack polynomials are given by,
\begin{eqnarray}
\label{jp}
&& J_{[1]}^{(\beta)}(p_k) = p_{1}\nonumber\;,\\
&& J_{[2]}^{(\beta)}(p_k) =  \dfrac{p_2 + \beta p_{1}^2}{\beta + 1}, \quad
 J_{[11]}^{(\beta)}(p_k) =  \dfrac{1}{2} \big( p_{1}^2 - p_2 \big) \;.
\end{eqnarray}

Schur functions, for $x$ an alphabet of cardinality $n$, are typically defined as a ratio of variables,
\begin{equation}\label{Eq_Schur-def}
\chi_{\lambda}(x)=
\frac{\det_{1\leqslant i,j\leqslant n}\big(x_i^{\l_j+n-j}\big)}{\Delta(x)},
\end{equation}
where $\Delta(x)$ is the Vandermonde product \eqref{Eq_Vandermonde}. Schur functions can be considered as a special case of Jack polynomials at $\b=1$.
\section{The Theorem and the Conjecture of Selberg integral}\label{coc}
Here we confirm the validity of conjecture \eqref{nschur}, 
which is actually equivalent to the theorem \eqref{Eq_nplusone}. We transform  \eqref{nschur} into
{\small
\begin{equation}\label{c1}
\begin{split}
&\left<\chi_{\bY^{(1)}}\left[\bx^{(1)}\right]...\chi_{\bY^{(r)}}\left[\bx^{(r)}-\bx^{(r-1)}-(\bv_1+...+\bv_{r-1})\right]...\chi_{\bY^{(n+1)}}\left[-\bx^{(n)}-(\bv_1+...+\bv_n)\right]\right>_{\bu_1,...,\bu_n,\bv}^{\bN_1,...,\bN_n}\\
=&\prod_{r=2}^{n+1}\bigg\{ (-1)^{|\bY^{(r)}|}\times\frac{\left[\bv_{r-1}+\bN_{r-1}-\bN_{r}\right]_{Y^{'(r)}}}{\left[\bN_{r-1}+\bN_{r}\right]_{Y^{'(r)}}}\times\prod_{1\leqslant i<j\leqslant\bN_{r}+\bN_{r-1}}\frac{(j-i+1)_{\bY^{'(r)}_i-\bY^{'(r)}_j}}{(j-i)_{\bY^{'(r)}_i-\bY^{'(r)}_j}}\bigg\}\times\\
&\times\prod_{1\leqslant i<j\leqslant\bN_{1}}\frac{(j-i+1)_{\bY^{(1)}_i-\bY^{(1)}_j}}{(j-i)_{\bY^{(1)}_i-\bY^{(1)}_j}}\times\\
&\times\prod_{1\leqslant r<s\leqslant n+1}\Bigg\{
\frac{\left[\bv_{s-1}+\bu_r+...+\bu_{s-1}-\bN_s+\bN_{s-1}\right]_{\bY^{'(s)}}}{\left[\bv_{s-1}-\bv_{r-1}+\bu_r+...+\bu_{s-1}-\bN_s+\bN_{s-1}-\bN_{r-1}\right]_{\bY^{'(s)}}}\times\\
&\qquad\qquad\qquad\times\frac{\left[-\bv_{r-1}+\bu_{r}+...+\bu_{s-1}+\bN_r-\bN_{r-1}\right]_{\bY^{(r)}}}{\left[\bv_{s-1}-\bv_{r-1}+\bu_{r}+...+\bu_{s-1}+\bN_r-\bN_{r-1}-\bN_s\right]_{\bY^{(r)}}}\times\\
&\times\prod_{i=1}^{\bN_{s-1}}\prod_{j=1}^{\bN_r}\frac{\bv_{s-1}-\bv_{r-1}+\bu_{r}+...+\bu_{s-1}+\bN_{s-1}+\bN_r-\bN_{r-1}-\bN_s+1-i-j}{\bv_{s-1}-\bv_{r-1}+\bu_{r}+...+\bu_{s-1}+\bN_{s-1}+\bN_r-\bN_{r-1}-\bN_s+1-i-j+\bY^{'(s)}_i+\bY^{(r)}_j}
\Bigg\},
\end{split}
\end{equation}
}
by setting $\bx^{(r)}=x^{(n+1-r)}$, $\bY^{(r)}=Y^{(n+2-r)}$, $\bv_r=v_{n+1-r}$, $\bu_r=u_{n+1-r}$ and $\bN_r=N_{n+1-r}$. We adopt a restriction on $v$, 
\begin{align*}
    \begin{split}
        \bv_{r}\equiv v_{n+1-r}=\left\{
        \begin{array}{ll}
             &  0,\quad otherwise\\
             &  v,\quad r=n.
        \end{array}
        \right.
    \end{split}
\end{align*}
Then we define $\bx^{(n+1)}=\bN_{n+1}=-\bv_n=-v$ And $A_{r,s}=\bu_r+...+\bu_{s-1}+\bN_r-\bN_{r-1}-\bN_s+\bN_{s-1}$ for $1\leqslant r<s \leqslant n+1$ then $A_{s,r}=-A_{r,s}$ otherwise. To simplify our notations, we would omit all bars over the letters. Thus \eqref{c1} becomes 
\begin{equation}\label{c2}
\begin{split}
&\left<\prod_{r=1}^{n+1}\chi_{Y^{(r)}}\left[x^{(r)}-x^{(r-1)}\right]\right>_{u_1,...,u_n,v}^{N_1,...,N_n}\\
=&\prod_{r=2}^{n}\bigg\{ (-1)^{|Y^{(r)}|}\times\frac{\left[N_{r-1}-N_{r}\right]_{Y^{'(r)}}}{\left[N_{r-1}+N_{r}\right]_{Y^{'(r)}}}\times\prod_{1\leqslant i<j\leqslant N_{r}+N_{r-1}}\frac{(j-i+1)_{Y^{'(r)}_i-Y^{'(r)}_j}}{(j-i)_{Y^{'(r)}_i-Y^{'(r)}_j}}\bigg\}\times\\
&(-1)^{|Y^{(n+1)}|}\times\frac{\left[N_{n}-N_{n+1}\right]_{Y^{'(n+1)}}}{\left[N_{n}\right]_{Y^{'(n+1)}}}\times\prod_{1\leqslant i<j\leqslant N_{n}}\frac{(j-i+1)_{Y^{'(n+1)}_i-Y^{'(n+1)}_j}}{(j-i)_{Y^{'(n+1)}_i-Y^{'(n+1)}_j}}\times\\
&\prod_{1\leqslant i<j\leqslant N_{1}}\frac{(j-i+1)_{Y^{(1)}_i-Y^{(1)}_j}}{(j-i)_{Y^{(1)}_i-Y^{(1)}_j}}\times\prod_{1\leqslant r<s\leqslant n+1}\Bigg\{
\frac{\left[A_{r,s}-N_r+N_{r-1}\right]_{Y^{'(s)}}}{\left[A_{r,s}-N_{r}\right]_{Y^{'(s)}}}\times\\
&\times\frac{\left[A_{r,s}+N_s-N_{s-1}\right]_{Y^{(r)}}}{\left[A_{r,s}-N_{s-1}\right]_{Y^{(r)}}}\times\prod_{i=1}^{N_{s-1}}\prod_{j=1}^{N_r}\frac{A_{r,s}+1-i-j}{A_{r,s}+1-i-j+Y^{'(s)}_i+Y^{(r)}_j}
\Bigg\}.
\end{split}
\end{equation}
The left hand sides of \eqref{c2} and \eqref{Eq_nplusone} are identical, which could be realized in the integral form directly. Thus we only need to confirm the equivalence of the right hand sides. A few results in \cite{Zhang:2011au} would be utilized are listed as follows,
\begin{equation}
    \frac{1}{[N]_Y}\prod_{1\leqslant i<j\leqslant N}\frac{(j-i+1)_{Y_i-Y_j}}{(j-i)_{Y_i-Y_j}}=\frac{1}{G_{Y,Y}(0)}
\end{equation}
\begin{equation}
    G_{Y^{'},Y^{'}}(x)=G_{Y,Y}(x), \qquad [x]_{Y^{'}}=(-1)^{|Y|}[-x]_Y=f_Y(x)
\end{equation}
\begin{equation}
    \prod_{i=1}^{N_1}\prod_{j=1}^{N_2}\frac{x+1-i-j}{x+1+Y^{'}_{i}+W_j-i-j}=\frac{(-1)^{|W|}[x-N_2]_{Y^{'}}[x-N_1]_{W}}{G_{Y,W}(x)G_{W,Y}(-x)}
\end{equation}
After applying these relations, the equation \eqref{c2} becomes
\begin{equation}
    \begin{split}
&\left<\prod_{r=1}^{n+1}\chi_{Y^{(r)}}\left[x^{(r)}-x^{(r-1)}\right]\right>_{u_1,...,u_n,v}^{N_1,...,N_n}\\
=&\prod_{r=1}^{n+1}\bigg\{ (-1)^{|Y^{(r)}|}\times\frac{f_{\Yar{r}}(N_{r-1}-N_r)}{G_{\Yar{r},\Yar{r}}(0)}\bigg\}\times\\
&\prod_{1\leqslant r<s\leqslant n+1}
\frac{f_{\Yar{s}}(A_{r,s}-N_r+N_{r-1})f_{\Yar{r}}(A_{s,r}-N_s+N_{s-1})}{G_{\Yar{s},\Yar{r}}(A_{r,s})G_{\Yar{r},\Yar{s}}(A_{s,r})},
\end{split}
\end{equation}
which is the same as \eqref{general} after translating the notations. Thus, we proved the identification of the Conjecture \eqref{nschur} and the Theorem \eqref{Eq_nplusone}.
%%%%%%%%%%%%%%%%%%%%%%%%%

\section{Proof of Lemmas}
\textbf{Lemma 1}
\begin{equation}\label{lemma1-s1}
    \prod_{1\leqslant i<j\leqslant L_Y}\frac{(Y_i-Y_j+\b(j-i))_{\b}}{(\b(j-i))_{\b}}\prod_{i=1}^{L_Y}\frac{1}{(\b(L_Y- i+1))_{Y_i}}
    =\frac{1}{G_{Y,Y}(0)}.
\end{equation}
\textbf{Proof}\\
Since $(x)_k=\frac{\Gamma(x+k)}{\Gamma(x)}$, we obtain
{\small
\begin{equation}
\begin{split}
    \frac{(Y_i-Y_j+\b(j-i))_{\b}}{(\b(j-i))_{\b}}=\frac{\Gamma((Y_i-Y_j)+\b(j-i+1)))}{\Gamma((Y_i-Y_j)+\b(j-i))}\times\frac{\Gamma(\b(j-i))}{\Gamma(\b(j-i+1))}=\frac{(\b(j-i+1))_{Y_i-Y_j}}{(\b(j-i))_{Y_i-Y_j}}.
\end{split}
\end{equation}
}
Thus the left hand side of \eqref{lemma1-s1} becomes
\begin{equation*}
    \begin{split}
        &\prod_{1\leqslant i<j\leqslant L_Y}\frac{(\b(j-i+1))_{Y_i-Y_j}}{(\b(j-i))_{Y_i-Y_j}}\prod_{i=1}^{L_Y}\frac{1}{(\b(L_Y-i+1))_{Y_i}}\\
        =&\prod_{1\leqslant i<j\leqslant L_Y}\frac{(\b(j-i+1))_{Y_i-Y_j}}{(\b(j-i))_{Y_i-Y_j}}\prod_{(i,j)\in Y}\frac{1}{\b(L_Y-i+1)+j-1}.
    \end{split}
\end{equation*}
Since Lemma 1 in \cite{Zhang:2011au} gives
\begin{align*}
    \prod_{1\leqslant i<j\leqslant L_Y}\frac{(\b(j-i+1))_{Y_i-Y_j}}{(\b(j-i))_{Y_i-Y_j}}=\frac{\prod_{(i,j)\in Y}{\b(L_Y-i+1)+j-1}}{G_{Y,Y}(0)},
\end{align*}
this helps us finish the proof.\\
\textbf{Lemma 2}
\begin{equation}
    \prod_{i=1}^{L_Y}(-z-\b i+\b)_{Y_i}=(-1)^{|Y|}f_{Y}(z)
\end{equation}
\textbf{Proof}\\
This proof only requires the definitions and canceling trivial terms.
\begin{equation}
\begin{split}
    &\prod_{i=1}^{L_Y}(-z-\b i+\b)_{Y_i}=\prod_{i=1}^{\ell_Y}(-z-\b i+\b)_{Y_i}\prod_{i=\ell_Y+1}^{L_Y}(-z-\b i+\b)_{0}=\prod_{i=1}^{\ell_Y}(-z-\b i+\b)_{Y_i}\\
    =&\prod_{i=1}^{\ell_Y}\prod_{j=1}^{Y_i}(-z-\b (i-1)+j-1)=\prod_{(i,j)\in Y}-\Big(z+\b(i-1)-(j-1)\Big)=(-1)^{|Y|}f_Y(z).
\end{split}
\end{equation}
The definition of $f_Y(z)$ is utilized in the last step.\\
\textbf{Lemma 3}
\begin{equation}\label{lemma3-p1}
    \begin{split}
     &\prod_{i=1}^{L_Y}\prod_{j=1}^{L_W}\frac{Y_i-W_j-x+j-i}{-x+j-i}
    \prod_{i=1}^{L_Y}\prod_{j=1}^{Y_i}\frac{1}{-x+L_W-i+j}
    \prod_{i=1}^{L_W}\prod_{j=1}^{W_i}\frac{1}{x+L_Y-i+j}\\
    =&\left.\frac{(-1)^{|Y|+|W|}}{G_{Y,W}(x)G_{W,Y}(-x)}\right|_{\beta=1}.
    \end{split}
\end{equation}
\textbf{Proof}\\
We use $\cL_n$ to represent the $n$th term of the left hand side of \eqref{lemma3-p1}. And $\cL_{i,j}$ for $j$th term of $\cL_i$. Firstly we decompose the products in $\cL_1$ into three parts.
\begin{align*}
    \cL_1=\prod_{i=1}^{\ell_Y}\prod_{j=1}^{\ell_W}\frac{Y_i-W_j-x+j-i}{-x+j-i}
    \prod_{i=\ell_Y+1}^{L_Y}\prod_{j=1}^{\ell_W}\frac{-W_j-x+j-i}{-x+j-i}
    \prod_{i=1}^{\ell_Y}\prod_{j=\ell_W+1}^{L_W}\frac{Y_i-x+j-i}{-x+j-i}.\\
\end{align*}
Notice that $\frac{Y_i-W_j-x+j-i}{-x+j-i}=1$ because when $i>\ell_Y$ and $j>\ell_W$, $Y_i=W_j=0$.
{\small
\begin{align*}
    \cL_{1,2}=&\prod_{i=\ell_Y+1}^{L_Y}\prod_{j=1}^{\ell_W}\frac{-W_j-x+j-i}{-x+j-i}
    =\prod_{i=\ell_Y+1}^{L_Y}\prod_{j=1}^{\ell_W}\prod_{k=1}^{W_j}\frac{-x+j-i-k}{-x+j-i-k+1}
    =\prod_{j=1}^{\ell_W}\prod_{k=1}^{W_j}\frac{-x+j-L_Y-k}{-x+j-\ell_Y-k}.\\
\end{align*}
}
Similarly,
\begin{equation*}
    \cL_{1,3}=\prod_{i=1}^{\ell_Y}\prod_{k=1}^{Y_i}\frac{-x-i+L_W+k}{-x-i+\ell_W+k}.
\end{equation*}
When $\ell_Y+1\leqslant i\leqslant L_Y$, we have $Y_i=0$. Thus
\begin{align*}
    \cL_2=&\prod_{i=1}^{L_Y}\prod_{j=1}^{Y_i}\frac{1}{-x+L_W-i+j}
    =\prod_{i=1}^{\ell_Y}\prod_{j=1}^{Y_i}\frac{1}{-x+L_W-i+j}.
\end{align*}
Similarly, for the third item
\begin{equation*}
    \cL_3=\prod_{i=1}^{\ell_W}\prod_{j=1}^{W_i}\frac{1}{x+L_Y-i+j}.
\end{equation*}
We notice that the numerators in $\cL_{1,2}$ and $\cL_{1,3}$ are respectively equal to the denominators in $\cL_3$ and $\cL_2$, out which can all be canceled.
\begin{align*}
    LHS=&\cL_1\cL_2\cL_3=\prod_{i=1}^{\ell_Y}\prod_{j=1}^{\ell_W}\frac{Y_i-W_j-x+j-i}{-x+j-i}(\cL_{1,2}\cL_3)(\cL_{1,3}\cL_2)\\
    =&\prod_{i=1}^{\ell_Y}\prod_{j=1}^{\ell_W}\frac{Y_i-W_j-x+j-i}{-x+j-i}\prod_{i=1}^{\ell_W}\prod_{j=1}^{W_i}\frac{1}{x-i+\ell_Y+j}
    \prod_{i=1}^{\ell_Y}\prod_{j=1}^{Y_i}\frac{1}{-x-i+\ell_W+j}.
\end{align*}

Thus, we need to prove that
{\small
\begin{equation}\label{lemma3-result}
    \prod_{i=1}^{\ell_Y}\prod_{j=1}^{\ell_W}\frac{Y_i-W_j-x+j-i}{-x+j-i}
    =\prod_{(i,j)\in Y}\frac{x+i-\ell_W-j}{x+W_i+Y_j^{'}-i-j+1}
    \prod_{(i,j)\in W}\frac{x-i+\ell_Y+j}{x-Y_i-W_j^{'}+i+j-1},
\end{equation}
}
in which some denominators are transposed to right. $G_{Y,W}(x)$ and $G_{W,Y}(-x)$ are rewritten by their definition. $(-1)^{|Y|}$ factor was absorbed by $\prod_{i=1}^{\ell_Y}\prod_{j=1}^{Y_i}\frac{1}{-x+L_W-i+j}$ and $(-1)^{|B|}$ by $G_{W,Y}(-x)$. Now we use mathematical induction to prove it.\par
Step 1. Proof for $W=\emptyset$.\par
In this case, $\ell_W=0$ and $W_i=0$ for any $i$.
\begin{align*}
    LHS=\prod_{i=1}^{\ell_Y}\prod_{j=1}^{0}\frac{Y_i-0-x+j-i}{-x+j-i}=1.
\end{align*}
which can be derived by equation \eqref{one}.
\begin{align}\label{lemma3-step1-rhs}
    RHS=\prod_{(i,j)\in Y}\frac{x+i-0-j}{x+0+Y_j^{'}-i-j+1}
    =\prod_{j=1}^{Y_1}\prod_{i=1}^{Y_j^{'}}\frac{x+i-j}{x+Y_j^{'}-i-j+1}=\prod_{j=1}^{Y_1}\prod_{i=1}^{Y_j^{'}}\frac{x+i-j}{x+i-j}=1.
\end{align}
The blocks in a Young diagram can be sorted by column,
\[\prod_{(i,j)\in Y}=\prod_{i=1}^{\ell_Y}\prod_{j}^{Y_i};\]
or by row,
\[\prod_{(i,j)\in Y}=\prod_{j=1}^{Y_1}\prod_{i=1}^{Y_j^{'}}.\]
In \eqref{lemma3-step1-rhs}, we separated them by the second way.\par
\begin{figure}[H]
    \centering
    \includegraphics[width=500pt,
    height=370pt]{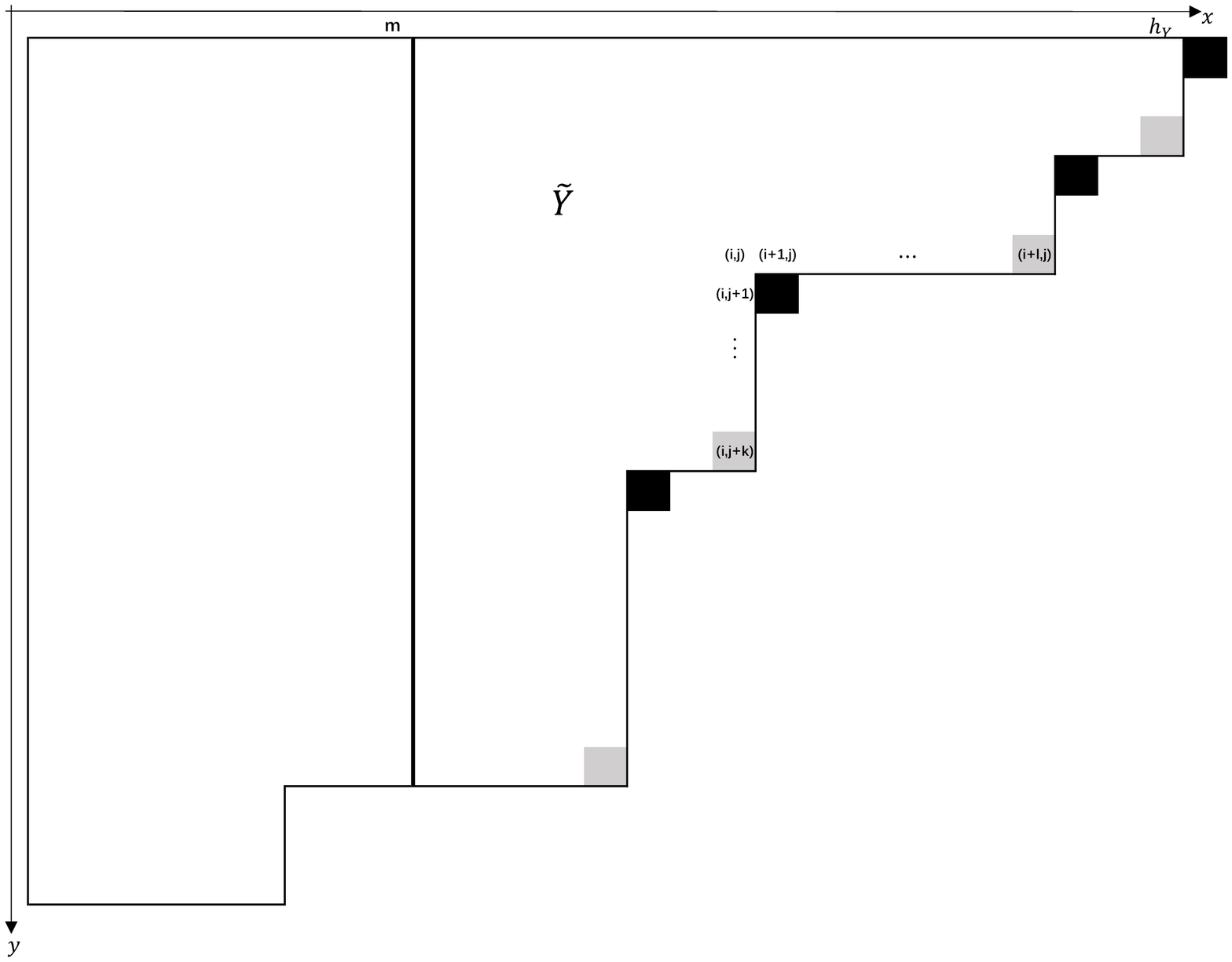}
    \caption{Grey boxes are in $R(\widetilde{Y})$ and black boxes are elements of $P(\widetilde{Y})$.}
    \label{lemma3-f1}
\end{figure}

Step 2. Induction for other cases. Suppose Lemma 3 is valid for W. Let us construct C which has only one cell different from W: $C_m=W_m+1, W_{W_m+1}^{'}=m-1, C_{W_m+1}^{'}=m$, 
with $m$ is the length of W. (Notice that the special case $W_m=0$ means $C_m$ starts from a new column, thus we can build any diagram from zero).\\
So we just need to prove that
\begin{align}\label{lemma3-p2}
    \prod_{i=1}^{\ell_Y}\prod_{j=1}^{\ell_C}\frac{Y_i-C_j-x+j-i}{-x+j-i}
    =\prod_{(i,j)\in Y}\frac{x+i-\ell_C-j}{x+C_i+Y_j^{'}-i-j+1}
    \prod_{(i,j)\in C}\frac{x-i+\ell_Y+j}{x-Y_i-C_j^{'}+i+j-1}.
\end{align}
We select extra items in equation \eqref{lemma3-p2} to conduct induction. The left hand side of \eqref{lemma3-p2} is
\begin{align*}
    LHS=\prod_{i=1}^{\ell_Y}\prod_{j=1}^{\ell_W}\frac{Y_i-W_j-x+j-i}{-x+j-i}\prod_{i=1}^{\ell_Y}\frac{Y_i-W_m-x+m-i-1}{Y_i-W_m-x+m-i}.
\end{align*}
We use $\cR_n$ to represent the $n$th term of right hand side of \eqref{lemma3-p2}.
{\small
\begin{align*}
    \cR_1=\prod_{(i,j)\in Y}\frac{x+i-\ell_C+j}{x+C_i+Y_j^{'}-i-j+1}
    =\prod_{(i,j)\in Y}\frac{x+i-\ell_W-j}{x+W_i+Y_j^{'}-i-j+1}\prod_{j=1}^{Y_m}\frac{x+Y_j^{'}-m-j+W_m+1}{x+Y_j^{'}-m-j+W_m+2}.
\end{align*}
}
Note that $\prod_{(i,j)\in C}f_{i,j}=\prod_{(i,j)\in W}f_{i,j}\times f_{m,W_m+1}$. $f_{i,j}$ is an arbitrary expression containing indexes $i$ and $j$.
\begin{align*}
    \cR_2=&\prod_{(i,j)\in C}\frac{x-i+\ell_Y+j}{x-Y_i-C_j^{'}+i+j-1}
    =\frac{x-m+\ell_Y+W_m+1}{x-Y_m+W_m}\prod_{(i,j)\in W}\frac{x-i+\ell_Y+j}{x-Y_i-C_j^{'}+i+j-1}\\
    =&\frac{x-m+\ell_Y+W_m+1}{x-Y_m+W_m}\prod_{(i,j)\in W}\frac{x-i+\ell_Y+j}{x-Y_i-W_j^{'}+i+j-1}
    \prod_{i=1}^{m-1}\frac{x-Y_i-m+i+W_m+1}{x-Y_i-m+i+W_m}.
\end{align*}
Lemma 3 holds for W,  we only need to prove that
\begin{equation}\label{lemma3-p3}
\begin{split}
    &\prod_{i=1}^{\ell_Y}\frac{x-Y_i+W_m-m+i+1}{x-Y_i+W_m-m+i}\\
    =&\prod_{j=1}^{Y_m}\frac{x+Y_j^{'}-m-j+W_m+1}{x+Y_j^{'}-m-j+W_m+2}\times
    \frac{x-m+\ell_Y+W_m+1}{x-Y_m+W_m}\prod_{i=1}^{m-1}\frac{x-Y_i-m+i+W_m+1}{x-Y_i-m+i+W_m}.
\end{split}
\end{equation}
LHS of equation \eqref{lemma3-p3} resembles the third term of RHS except the times of product. If $m\geqslant \ell_Y$, the LHS was canceled wholly. And $Y_i$=0 for all $i>\ell_Y$, leading the first item of RHS disappeared. Thus, 
\begin{align*}
    RHS\eqref{lemma3-p3}=&\frac{x-m+\ell_Y+W_m+1}{x+W_m}\prod_{i=\ell_Y+1}^{m-1}\frac{x-m+i+W_m+1}{x-m+i+W_m}\\
    =&\frac{x-m+\ell_Y+W_m+1}{x+W_m}\frac{x-m+m-1+W_m+1}{x-m+\ell_Y+W_m+1}=1.
\end{align*}
In the case of $\ell_Y\geqslant m$. For conciseness, we define a parameter
\begin{align*}
    a&=x+W_m-m+1.
\end{align*}
The equation which we need to prove becomes
\begin{equation}
    \prod_{i=m}^{\ell_Y}\frac{a-Y_i+i}{a-Y_i+i-1}
    =\frac{a+\ell_Y}{a+m-Y_m-1}\prod_{j=1}^{Y_m}\frac{a+Y_j^{'}-j}{a+Y_j^{'}-j+1}.
\end{equation}
This equation can be transformed to,
\begin{equation}\label{lemma3-p4}
    \prod_{i=m+1}^{\ell_Y}\frac{a-Y_i+i}{a-Y_i+i-1}
    =\frac{a+\ell_Y}{a+m-Y_m}\prod_{j=1}^{Y_m}\frac{a+Y_j^{'}-j}{a+Y_j^{'}-j+1}.
\end{equation}
We define a function for a block in a Young diagram.
\begin{equation}
    F(\square)=a+x(\square)-y(\square).
\end{equation}
where $x(\square)$ and $y(\square)$ respectively mean the column number and row number of a block. If a block is represented by its position $(x,y)$, we define
\begin{equation}
    F(x,y)=a+x-y.
\end{equation}
For further calculation, we give a notation $\widetilde{Y}$, which contains the blocks whose abscissas are greater than $m$. Numerator in left hand side of \eqref{lemma3-p4} is the product of all blocks reside at the endpoints of columns in $\widetilde{Y}$. Denominator is the product of the left adjoining squares of these blocks. Now we give a proposition:
\be\label{lemma3-left}
\prod_{i=m+1}^{\ell_Y}\frac{a-Y_i+i}{a-Y_i+i-1}=\frac{F(\ell_Y+1,1)}{F(m,Y_m)}\times\frac{\prod_{\square\in R(\widetilde{Y})}F(\square)}{\prod_{\square\in P(\widetilde{Y})}F(\square)}.
\ee
where $R(\widetilde{Y})$ means the blocks which could be removed from $\widetilde{Y}$. $P(\widetilde{Y})$ means the position where a cell could be added to $\widetilde{Y}$.
We take a sequence of endpoints of columns having the same length for example to confirm the validity (See figure. \ref{lemma3-f1}). In this case, we assume $Y_{m}=Y_{m+1}$. Another case can be proved by the same way.
\be
\begin{split}
    \frac{F(i+1,j)}{F(i,j)}\frac{F(i+2,j)}{F(i+1,j)}\cdots\frac{F(i+l,j)}{F(i+l-1,j)}=\frac{F(i+l,j)}{F(i,j)}=\frac{F(i+l,j)}{F(i+1,j+1)}.
\end{split}
\ee
The block $(i+1,j+1)$ is an element of $P(\widetilde{Y})$ and the block $(i+l,j)$ is an element of $R(\widetilde{Y})$. The same process can be applied to every sequence of columns' endpoints. However, block $(\ell_Y+1,1)$ is in $P(\widetilde{Y})$ but $F(\ell_Y+1,1)$ is not in LHS of \eqref{lemma3-left}. And $F(m,Y_m)$ is in LHS of \eqref{lemma3-left} but block $(m,Y_m)$ is not in $P(\widetilde{Y})$. Thus we need multiple $\frac{\prod_{\square\in R(\widetilde{Y})}F(\square)}{\prod_{\square\in P(\widetilde{Y})}F(\square)}$ by $\frac{F(\ell_Y+1,1)}{F(m,Y_m)}$ to get LHS of \eqref{lemma3-left}.

The individual fraction in right hand side of \eqref{lemma3-p4} is $\frac{F(\ell_Y+1,1)}{F(m,Y_m)}$. And the product is concerned with the endpoints of rows in $\widetilde{Y}$, equal to $\frac{\prod_{\square\in R(\widetilde{Y})}F(\square)}{\prod_{\square\in P(\widetilde{Y})}F(\square)}$. We take a sequence to embody the validity(see figure.\ref{lemma3-f1}).
\be
\frac{F(i,j+1)}{F(i,j)}\frac{F(i,j+2)}{F(i,j+1)}\cdots\frac{F(i,j+m)}{F(i,j+m-1)}=\frac{F(i,j+m)}{F(i,j)}=\frac{F(i,j+m)}{F(i+1,j+1)}.
\ee
The block $(i+1,j+1)$ is an element of $R(\widetilde{Y})$ and the block $(i,j+m)$ is an element of $P(\widetilde{Y})$. The same process can be applied to every sequence of rows' endpoints to get complete $\frac{\prod_{\square\in R(\widetilde{Y})}F(\square)}{\prod_{\square\in P(\widetilde{Y})}F(\square)}$.
Thus we finish the proof of \eqref{lemma3-p4},  resulting in the proof of Lemma 3.
%%%%%%%%%%%%%%%%%%%%%%%%%
\section{Proof for equation \eqref{x-part}}\label{app-x}
Things are a little more complicated for equation \eqref{x-part}. We need set the inverse terms coincide, i.e. $z^{(r)}=x^{(n+1-r)}$, $N_r=N_{(n+1-r)+}$, $W^{(r)}=Y^{(n+2-r)}$, $u_r=u_{(n+1-r)+}$ for any $0\leqslant r\leqslant n+1$ and $v=v_+$. Firstly we confirm the equivalence of these two averages by the definition of Selberg integral \eqref{Eq_I-def}.
\begin{equation}
    \begin{split}
    I_{N_1,...,N_n}^{A_n}\left(\prod_{r=1}^{n+1}\chi_{W^{(r)}}\left[z^{(r)}-z^{(r-1)}\right];u_1,...,u_n,v\right)=&\\
    \frac{1}{(2\pi\iup)^{N_1+\dots+N_n}} \Int_{C^{N_1,\dots,N_n}}\prod_{r=1}^{n+1}\chi_{W^{(r)}}\left[z^{(r)}-z^{(r-1)}\right]&\prod_{r=1}^{n}\left[|\Delta(z^{(r)})|^2\prod_{i=1}^{N_r}(z^{(r)}_{i})^{u_r}\right]\times\\
    \prod_{i=1}^{N_n}(1-z^{(n)}_i)^{v}&\prod_{r=1}^{n-1}|\Delta(z^{(r)},z^{(r+1)})|^{-1}\prod_{r=1}^{n}dz^{(r)}.
    \end{split}
\end{equation}
And the integral appears in AGT correspondence,
\begin{equation}
    \begin{split}
        I_{N_{1+},...,N_{n+}}^{A_n}\left(\prod_{r=1}^{n+1}\chi_{Y^{(r)}}\left[x^{(r-1)}-x^{(r)}\right];u_{1+},...,u_{n+},v_+  
                   \right)=&\\
        \frac{1}{(2\pi\iup)^{N_{1+}+\dots+N_{n+}}} \Int_{C^{N_{1+},\dots,N_{n+}}}\prod_{r=1}^{n+1}\chi_{Y^{(r)}}\left[x^{(r-1)}-x^{(r)}\right]&\prod_{r=1}^{n}\left[|\Delta(x^{(r)})|^2
        \prod_{i=1}^{N_{r+}}(x^{(r)}_{i})^{u_{r+}}\right]\times\\
        \prod_{i=1}^{N_{1+}}(1-x^{(1)}_i)^{v_+}&\prod_{r=1}^{n-1}|\Delta(x^{(r)},x^{(r+1)})|^{-1}\prod_{r=1}^{n}dx^{(r)}.
    \end{split}
\end{equation}
We would check one term to explain why they are equal. The product of Schur polynomials,
\begin{equation}\label{first}
    \begin{split}
        &\prod_{r=1}^{n+1}\chi_{W^{(r)}}\left[z^{(r)}-z^{(r-1)}\right]=\prod_{r=1}^{n+1}\chi_{Y^{(n+2-r)}}\left[x^{(n+1-r)}-x^{(n+1-(r-1))}\right]\\
        =&\prod_{r=1}^{n+1}\chi_{Y^{(n+2-r)}}\left[x^{((n+2-r)-1)}-x^{(n+2-r)}\right]=\prod_{r=1}^{n+1}\chi_{Y^{(r)}}\left[x^{(r-1)}-x^{(r)}\right].
    \end{split}
\end{equation}
$z^{(r)}=x^{(n+1-r)}$ and $W^{(r)}=Y^{(n+2-r)}$ are used in the first step. In the last step, the products of items from $n+2-r=n+1$ to $n+2-r=1$ are translated to the products of items from $r=1$ to $r=n+1$.
All items in two integrals could be equated respectively by the similar procedure. Thus,
{\small
\begin{equation}
    \begin{split}
    &\left<\prod_{r=1}^{n+1}\chi_{Y^{(r)}}\left[x^{(r-1)}-x^{(r)}\right]\right>_+=\left<\prod_{r=1}^{n+1}\chi_{W^{(r)}}\left[z^{(r)}-z^{(r-1)}\right]\right>_{u_1,...,u_n,v}^{N_1,...,N_n}\\
        &=\prod_{r=1}^{n+1}(-1)^{|W^{(r)}|}\prod_{r,s=1}^{n+1}\frac{f_{W^{(r)}}(-(\sum_{i=r}^{s-1}u_i-\sum_{i=s}^{r-1}u_i+N_r-N_{r-1}))}{G_{W^{(r)},W^{(s)}}(-(\sum_{i=r}^{s-1}u_i-\sum_{i=s}^{r-1}u_i+N_r-N_{r-1}-N_s+N_{s-1}))}\\
        &=\prod_{r=1}^{n+1}(-1)^{|Y^{(r)}|}\times  \prod_{r,s=1}^{n+1}\frac{f_{Y^{(r)}}(-\sum_{i=s}^{r-1}u_{i+}+\sum_{i=r}^{s-1}u_{i+}-N_{(r-1)+}+N_{r+})}{G_{Y^{(r)},Y^{(s)}}(-\sum_{i=s}^{r-1}u_{i+}+\sum_{i=r}^{s-1}u_{i+}-N_{(r-1)+}+N_{r+}+N_{(s-1)+}-N_{s+})}.
    \end{split}
\end{equation}
}
In the first step, we changed our variables to use the integral formula. And we finish the integral and return to our initial variables in the second and last steps.

Now we give the relations of parameters to finalize our proof.
\begin{equation}
    \begin{split}
        \mu_{s}+a_r&=\sum_{i=r}^{s-1}u_{i+}-\sum_{i=s}^{r-1}u_{i+}+N_{r+}-N_{(r-1)+},\\
        a_r-a_s&=\sum_{i=r}^{s-1}u_{i+}-\sum_{i=s}^{r-1}u_{i+}+N_{r+}-N_{(r-1)+}-N_{s+}+N_{(s-1)+}.
    \end{split}
\end{equation}
$r$ and $s$ run from $1$ to $n+1$. After applying these
parameters,  equation \eqref{x-part} is obtained.

\bibliographystyle{unsrt}
\bibliography{AGT_proof}

\end{document}